\documentclass[12pt, a4paper, twoside]{article}
\pdfoutput=1
\usepackage[a4paper, left=3cm,right=2cm]{geometry}
\usepackage{hyperref}
\usepackage{amsfonts}
\usepackage{amsmath}
\usepackage{empheq}
\usepackage{setspace}
\usepackage{color}
\usepackage{slashed}
 \usepackage{hyperref}
 \usepackage{mathrsfs}
  
\usepackage[utf8,applemac]{inputenc}
\usepackage{cite}
\usepackage{tikz}
\usepackage{graphicx}% Include figure files
\graphicspath{{figure/}}
\usepackage{dcolumn}% Align table columns on decimal point
\usepackage{bm}% bold math

\newcommand{\bea}{\begin{eqnarray}}
\newcommand{\eea}{\end{eqnarray}}
\newcommand{\be}{\begin{equation}}
\newcommand{\ee}{\end{equation}}

\def\p{\partial}

\title{The Mass of Kerr-Newman Black Holes in an external magnetic field}

\begin{document}

%\maketitle
%\abstract{[...]}

\setcounter{tocdepth}{2}

\begin{titlepage}

\begin{flushright}\vspace{-3cm}
CECS-PHY-16/01 \\
UAI-PHY-16/03\\
\vspace{0.3cm}
{\small
%{\tt arXiv:yymm.nnnn} \\
\today }\end{flushright}
\vspace{0.9cm}

\begin{center}
{{ \LARGE{\bf{Mass of Kerr-Newman Black Holes \\ \vspace{3mm} in an external magnetic field\\ }}}} \vspace{8mm}

\centerline{\large{M. Astorino$^{a,b, }$\footnote{e-mail: marco.astorino@gmail.com}, G. Comp\`{e}re$^{c, }$\footnote{e-mail: gcompere@ulb.ac.be}, R. Oliveri$^{c, }$\footnote{e-mail: roliveri@ulb.ac.be}, N. Vandevoorde$^{c, }$\footnote{e-mail: nvdevoor@ulb.ac.be},}}

\vspace{5mm}
\textit{
$^{a}${Centro de Estudios Cient\'{\i}ficos (CECs),\\ Valdivia, Chile\\}}

\bigskip\medskip
\textit{
$^{b}${Departamento de Ciencias, Facultad de Artes Liberales,\\
% Facultad de Ingenier\'{\i}a y Ciencias\\
Univesidad Adolfo Iba\~{n}ez, \\ 
Av. Padre Hurtado 750, Vi\~{n}a del Mar, Chile}
}

\normalsize
\bigskip\medskip
\textit{
$^{c}${Universit\'{e} Libre de Bruxelles and International Solvay Institutes\\
CP 231, B-1050 Brussels, Belgium}
}

%\vfil
%\pacs{04.70.Dy}

\vspace{5mm}

\begin{abstract}
\noindent 

The explicit solution for a Kerr-Newman black hole immersed in an external magnetic field, sometimes called the Melvin-Kerr-Newman black hole, has been derived by Ernst and Wild in 1976. In this paper, we clarify the first law and Smarr formula for black holes in a magnetic field. We then define the unique mass which is integrable and reduces to the Kerr-Newman mass in the absence of magnetic field. This defines the thermodynamic potentials of the black hole. Quite strikingly, the mass coincides with the standard Christodoulou-Ruffini mass of a black hole as a function of the entropy, angular momentum and electric charge.

\end{abstract}

%\pacs{04.65.+e,04.70.-s,11.30.-j,12.10.-g}

\end{center}
%%%%%%%%%%%%%%%%%%%%%%%%%%%%%%%%%%%%%%%%%%%%%%%%%%%%%%%%%%%%%%%%%%%%%%%%%%%%%%%%%%%%%%%%

\end{titlepage}

\newpage
\tableofcontents

\section{Introduction and summary}

It is well known that, at the center of galaxies, strong axial magnetic fields are generated by infalling matter in the accretion disk of the black hole galactic nucleus, as is measured in our Milky Way \cite{Eatough:2013nva}. On the other hand, few exact analytical models, which describe black holes interacting with external magnetic fields, are known. 

In the framework of general relativity coupled to the Maxwell field, Ernst and Wild \cite{ernst1976kerr}, thanks to the solution-generating techniques developed in \cite{Ernst:1967wx,Ernst:1967by},  were able to build an exact and regular solution for a magnetized black hole, which generalizes the Kerr-Newman one. The Ernst-Wild metric is a stationary and axisymmetric solution to the Einstein-Maxwell equations, describing a Kerr-Newman black hole embedded in a backreacting external magnetic field. In the weak magnetic field approximation, the gauge field is exactly a linear combination of the two covariant components of the Killing vectors of the Kerr-Newman metric \cite{Wald:1974np}. In the strong magnetic regime, the interaction with the electromagnetic field affects the conserved charges of the black hole, such as the angular momentum, the electric charge and the mass. Moreover, the presence of the electromagnetic field deforms the asymptotic region of the solution: the metric approaches, for large radial distances, the Melvin magnetic universe \cite{melvin1964pure}. It is then often referred to as the Melvin-Kerr-Newman black hole.\footnote{Note that the electromagnetic field does not   coincides exactly with the one of the Melvin geon \cite{hiscock1981black}, but only up to a electromagnetic duality transformation \cite{Gibbons:2013yq}.} It has been noted that unphysical pathologies occur at large distances: for example, ergoregions form around the poles \cite{Gibbons:2013yq}. The astrophysical role of this solution is, therefore, unclear. The solution has, however, other applications. In the extremal limit, it can serve as a useful additional toy model for the so-called Kerr/CFT correspondence \cite{Guica:2008mu} since the near-horizon geometry of the extremal Melvin-Kerr-Newman black hole exactly matches with the one of the extremal Kerr-Newman black hole \cite{Astorino:2015naa,Bicak:2015lxa} (see also \cite{Siahaan:2015xia}). It, therefore, provides an alternative geometry which extends the near-horizon region of the extremal Kerr-Newman black hole \cite{Bardeen:1999px} outside of the near-horizon regime.

While several relevant features of this solution have been disclosed since its publication, some aspects of the Ernst-Wild solution, concerning the energy and thermodynamics, still remain controversial due to the unconventional asymptotic region. In the last years, several attempts have been made in that direction by using different theoretical methods, but still there is no general agreement; see, for instance, \cite{dokuchaev1987black,aliev1988effective,aliev1989magnetized,karas1991interpretation,Gibbons:2013dna,Astorino:2015naa,Booth:2015nwa2}. 

The scope of this paper is to address this open issue. We will define the mass of these magnetized black holes thanks to the methods developed in \cite{Barnich:2001jy,Barnich:2003xg,Barnich:2007bf}, which we will refer to as the canonical integrability methods. These methods are especially well adapted to the problem at hand since the asymptotic region has no role to play. This method was already applied previously to black holes with unusual asymptotics \cite{Barnich:2005kq,Banados:2005da,Compere:2007vx}. The lack of known boundary conditions for Melvin universes is, therefore, not an obstacle for defining the mass. The procedure amounts to first define the infinitesimal change of mass due to infinitesimal changes of parameters by considering the integral of a uniquely defined surface charge form on an arbitrary sphere surrounding the black hole. One then write the conditions for the existence of a finite mass known as the integrability conditions. The definition of the mass amounts to solving these integrability conditions. The rest of the thermodynamic analysis is standard and directly follows from the definition of the mass. 

Before starting our analysis, let us first summarize the main results of the paper:
\begin{itemize}
\item[1.] We first clarify the form of the first law of black hole mechanics in the presence of an external magnetic field $B$ in order to show that there is no $\delta B$ term:
\bea
\delta \mathcal{M} = T \delta S + \Omega \delta J + \Phi \delta Q.\label{fl}
\eea
Here, $\mathcal{M}$ is the canonical mass and all parameters are varied, including the magnetic field. Our proof is standard and only uses the methods of Bardeen-Hawking-Carter \cite{Bardeen:1973gs} and Iyer-Wald \cite{Iyer:1994ys}. We will also show that the Smarr formula without $B$ term directly follows from the first law. 

\item[2.] We show that the integrability condition for the mass leads to a single ``master equation'' for the mass as a functional of the four black hole parameters ($m, a, q, B$). This master equation turns out to be a partial differential equation in the four black hole parameters.

\item[3.] We then show that the master equation admits a unique solution, provided the boundary condition that the mass is equal to the Kerr-Newman mass in the absence of magnetic field. Our procedure, therefore, provides us with the unique definition of the canonical mass for the Melvin-Kerr-Newman spacetimes. 
%In terms of the mass, rotation, electric charge and magnetic field parameters $(m,a,q,B)$ it is given by
%\begin{align} 
%	\mathcal{M}^{2}&= m^{2} + 2 j q B + \Bigl(2 j^{2} + \frac{3}{2}m^{2}q^{2}-q^{4}\Bigr)B^{2} + \\
%	\qquad &+ j q\Bigl(2 m^{2} - \frac{3}{2} q^{2}\Bigr)B^{3}+\Bigl(j^{2}m^{2} - \frac{1}{2} j^{2} q^{2} + \frac{1}{16} m^{2} q^{4}\Bigr)B^{4}. 		\nonumber
%	\end{align}
It turns out that the formula for the mass exactly coincides with the one obtained in \cite{Booth:2015nwa2} by means of the isolated-horizon approach (the mass did not however appeared in the published version \cite{Booth:2015nwa3}). Consistently with the first law \eqref{fl}, the mass can be expressed as a function of the black hole entropy $S$, angular momentum $J$ and electric charge $Q$ without any explicit appearance of the magnetic field $B$. Quite strikingly, the resulting mass formula $M(S,Q,J)$ exactly coincides with the Christodoulou-Ruffini mass as defined for the Kerr-Newman black hole in \cite{Christodoulou:1972kt}. 

As a cross-check, one can compare the mass in the extremal limit with the one obtained in \cite{Astorino:2015naa} and \cite{Bicak:2015lxa} by a totally different procedure. In \cite{Astorino:2015naa,Bicak:2015lxa}, the authors first matched the parameters of the identical near-horizon geometries of the extremal Kerr-Newman black hole and the extremal Melvin-Kerr-Newman black hole. Second they inferred the mass of the extremal Melvin-Kerr-Newman black hole from the mass of the extremal Kerr-Newman black hole. It turns out that the extremal limit of the mass that we found exactly matches with the results of \cite{Astorino:2015naa} and \cite{Bicak:2015lxa}.
	
It is quite remarkable that Melvin-Kerr-Newman black holes share the same thermodynamic relations as the Kerr-Newman black hole. It was known that the near-horizon geometry of both Melvin-Kerr-Newman and Kerr-Newman agree in the extremal limit. Here, we find a relationship away from extremality. \\
Since this characteristic behavior occurs also for other deformations of the Kerr-Newman metric, for instance in the presence of acceleration \cite{acc-mkn}, we think it might be a general feature for regular, axisymmetric and stationary electrovacuum black holes.  
	
\item[4.] Knowing the canonical generator associated with the mass, one can find the canonical frame such that the mass is associated with time translations by a unique change of coordinates and gauge at infinity. We provide the explicit frame change from the coordinates and gauge defined in \cite{Gibbons:2013yq} to the canonical frame.

\item[5.] Finally, for completeness, we also consider the alternative situation where the magnetic field is considered as an external source with no dynamics. We show that one integrability constraint is lifted, which allows for a one-function family of consistent thermodynamics. The function left arbitrary by the integrability method is precisely the alternative mass $\widetilde{\mathcal M} $. Since the canonical mass is no longer integrable under variations of the magnetic field $B$, the first law \eqref{fl} reads instead
\bea
\slashed{\delta} \mathcal{M} = T \delta S + \Omega \delta J + \Phi \delta Q.
\eea
One can then identify the left-hand side as $\slashed{\delta} \mathcal{M} = \delta \widetilde{\mathcal M} + \mu \delta B$ in terms of the variation of the alternative mass and the variation of the magnetic field source $B$ times its conjugate chemical potential, the magnetic dipole moment. One such example is the thermodynamics studied in \cite{Gibbons:2013yq}, where the mass was fixed from other considerations. 
	
\end{itemize}

In the following, we shall use the Melvin-Kerr-Newman metric and gauge field as written in \cite{Gibbons:2013yq} except for three minor adaptations that we explain in Appendix A. The choice of coordinate frame and gauge in \cite{Gibbons:2013yq} will turn out to be irrelevant for our purposes since all our derivation will be frame and gauge independent as we will show. Some well-established thermodynamical quantities at the horizon of the Melvin-Kerr-Newman black hole are also recalled in Appendix A. The rest of the paper is organized as follows. In Sec. \ref{sec 2}, we prove the first law of black hole thermodynamics in the presence of an external magnetic field. In Sec.  \ref{sec 3}, we implement the canonical integrability methods and find the unique mass.  In Sec.  \ref{sec 4} we discuss the properties of the mass and derive the remaining thermodynamical potentials. We also define the canonical frame. We finally discuss the alternative thermodynamics that arises when the magnetic field is treated as a nondynamical field. 

\section{First law with an external magnetic field} \label{sec 2}

The main aim of this section is to prove that the first law of black hole mechanics in the presence of external magnetic field $B$ contains no $\delta B$ term contrary to some earlier claims. This conclusion is a direct consequence of the geometrical derivation of the first law.  The original proof of the first law was done for asymptotically flat spacetimes by Bardeen-Carter-Hawking \cite{Bardeen:1973gs}. However, one readily generalize it to any asymptotics using the universal definition of infinitesimal charge associated with canonical generators as done by Iyer-Wald in pure gravity using covariant phase space methods \cite{Iyer:1994ys} (this derivation was extended with the Maxwell field included in \cite{Rogatko:2002eu,Gao:2003ys}). 

We want to emphasize that there are no subtleties related to the presence of the external magnetic field. Indeed, the metric is smooth and the gauge field is regular outside the black hole. In particular, in the following derivation, we have assumed that no magnetic monopole is present and we have used a gauge potential regular at the poles. In the presence of magnetic monopoles or dipoles where the gauge field is singular, subtleties in the geometrical derivation of the first law are present and lead to an additional term \cite{Copsey:2005se}.

Before starting the proof, we would like to mention the notation and the nomenclature of the formalism used in the following (see e.g. \cite{Compere:2006my} for a review).
In the context of Einstein-Maxwell theory, the generalized Killing equations for the fields $g_{\mu\nu}$ and $A_{\mu}$ are
\bea
\mathcal{L}_{\xi}g_{\mu\nu}=0, \qquad \mathcal{L}_{\xi}A_\mu+ \p_\mu \lambda=0,
\eea
where $\xi = \xi^\mu \p_\mu$ is a Killing vector field and $\lambda$ is a real constant. We call symmetry parameter the pair $(\xi, \lambda)$, which is solution to the generalized Killing equations. It can be shown that a surface charge $k_{(\xi, \lambda)}[\delta g, \delta A; g, A]$ is associated with the symmetry parameter $(\xi, \lambda)$ and it is uniquely fixed as a functional of the Lagrangian, up to an irrelevant total derivative \cite{Barnich:2001jy}. The surface charges $k_{(\xi, \lambda)}$ are spacetime $2$-forms and $1$-forms in the field space. The explicit formula for the surface charge that we will use in this paper can be found in (4.22) of\cite{Compere:2009dp} (where one sets the scalar field to zero, $\chi = 0$, $h_{IJ} = 0$ and $k_{IJ}=1$). 

 The total conserved charge $\mathcal{Q}_{(\xi, \lambda)}$ associated to $(\xi, \lambda)$ is defined by
\bea
\mathcal{Q}_{(\xi, \lambda)}[g,A ;  \bar{g}, \bar{A}]= \int_S \int_{\bar{g}}^{g} \int_{\bar{A}}^{A}  k_{(\xi, \lambda)}[\delta g', \delta A'; g', A'],\label{defQ}
\eea
where $S$ is a spacetime two-surface. Such a definition is meaningful provided that the surface charge does not depend on the path in the field space. This requirement is called the integrability condition and reads as 
\bea
\int_S \delta_1 k_{(\xi, \lambda)}[\delta_2 g, \delta_2 A; g', A'] - (1 \leftrightarrow 2) = 0. \label{intcond}
\eea
Since the surface charge is closed on-shell \cite{Barnich:2001jy}, the charge \eqref{defQ} is conserved under any deformation of the surface. It is, therefore, radius and time independent. 

The first law of black hole mechanics is essentially a consequence of the following conservation law:
\bea \label{cons law}
\int_H k_{(\xi,0)} = \int_S k_{(\xi,0)},
\eea
where $\xi=\p_t + \Omega_H \p_\phi$ is the Killing generator of the black hole horizon and $\Omega_H$ is the angular velocity of the horizon.
The equivalence is between the surface charge integrated over the spacelike section of the black hole horizon $H$ and over the sphere at infinity $S$, even though any sphere enclosing the horizon is equally valid for the argument.

At the horizon, the standard derivation \cite{Bardeen:1973gs,Iyer:1994ys} leads to
\bea
\int_H k_{(\xi,0)} = T_H \delta S + \Phi_H \delta Q,
\eea
where the chemical potential $T_H$, associated to the entropy $S$, is the Hawking temperature. The Coulomb electrostatic potential at the horizon is defined as $\Phi_H= - A_{\mu}\xi^{\mu}|_H$ and $Q$ is the electric charge.

In order to develop the right-hand side of \eqref{cons law}, it is necessary to identify which is the canonical symmetry parameter associated with the energy. In general, it is not $(\p_t, 0)$. In other words, one needs to consider the most general symmetry generator $(\alpha \chi, \alpha \Phi)$, where $\chi= \p_t + \Omega \p_\phi$ is a Killing vector field and $\Phi$ is a gauge transformation parameter. Here, $\alpha$, $\Omega$ and $\Phi$ are spacetime constants but phase space functions: they are allowed to depend upon the black hole metric parameters. We will prove in Sec. \ref{sec 3} that the mass exists and is uniquely defined as the conserved charge associated with such a general symmetry generator $ \alpha_{int} (\p_t + \Omega_{int} \p_\phi, \Phi_{int} )$. The three unknown functions $\alpha_{int}$, $\Omega_{int}$ and $ \Phi_{int}$ will be uniquely fixed by the integrability conditions. In the rest of this section we shall not need the explicit values of these functions. 

In order to prove the first law, we can use the linearity of the definition of symmetry parameters and reorganize the symmetry parameter as follows:
\bea \label{symmetry parameter}
\alpha(\p_t + \Omega_H \p_\phi, 0) =  \alpha(\p_t + \Omega_{int} \p_\phi, \Phi_{int}) + \alpha((\Omega_{H} - \Omega_{int}) \p_\phi, -\Phi_{int}).
\eea
Since the surface charge is linear in the symmetry parameter, Eq. \eqref{cons law} becomes
\bea \label{cons law2}
\int_H k_{(\alpha \xi, 0)}=\int_S k_{\alpha(\chi, \Phi_{int})} + \int_S k_{\alpha((\Omega_{H} - \Omega_{int}) \p_\phi, - \Phi_{int})}.
\eea
Using the following definitions,
\bea
\delta\mathcal{M}=\int_S k_{\alpha(\chi, \Phi_{int})}, \qquad \delta J = \int_S k_{(-\p_\phi, 0)}, \qquad \delta Q = \int_S k_{(0, -1)},
\eea
Eq. \eqref{cons law2} reads
\bea
\alpha(T_H \delta S + \Phi_H \delta Q)=\delta\mathcal{M} - \alpha (\Omega_{H} - \Omega_{int})\delta J + \alpha\Phi_{int} \delta Q,
\eea
from which we obtain
\bea
\label{first law}
\boxed{
\delta \mathcal{M} = \alpha \Bigl( T_H \delta S + (\Omega_{H} - \Omega_{int} ) \delta J + (\Phi_{H} - \Phi_{int}) \delta Q \Bigr)}
\eea
which is the first law of black hole mechanics. Under a change of frame, $t \rightarrow \tilde t = \Delta_t \, t $, $\phi \rightarrow \tilde \phi = \phi + \Delta  \Omega\, t$ and a change of gauge $A \rightarrow \tilde A = A + d\lambda$ with $\lambda = -  (\Delta A_t ) t$ (leading to the total transformation $A_t \rightarrow \tilde A_{\tilde t} = \Delta_t^{-1} (A_t - \Delta \Omega A_\phi- \Delta A_t )$) the potentials appearing in the first law transform as 
\bea
T_H &\rightarrow & T_H \Delta_t^{-1}, \qquad \qquad\qquad\quad\,  \alpha\;\;\rightarrow\;\; \alpha  \Delta_t , \nonumber\\
\Omega_{H} &\rightarrow &\left( \Omega_H +  \Delta  \Omega \right) \Delta_t^{-1} , \qquad\; \Omega_{int}\;\;\rightarrow\;\ \left( \Omega_{int} +   \Delta  \Omega \right) \Delta_t^{-1} ,\label{cc}\\
\Phi_{H} &\rightarrow &\left( \Phi_{H} + \Delta A_t \right) \Delta_t^{-1} , \qquad \Phi_{int}\;\;\rightarrow\;\ \left( \Phi_{int} + \Delta A_t\right) \Delta_t^{-1} .\nonumber
\eea
Here $\Delta \Omega$, $\Delta_t$ and $ \Delta A_t$ are constants. The last transformation law follows from the fact that the surface charge $k_{( \p_t, 0)}$ transforms under the large gauge transformation $A \rightarrow A+d\lambda$ generated by $\lambda = \Delta A_t \, t$ as $\oint_S k_{( \p_t, 0)} \rightarrow \oint_S k_{( \p_t, 0)} - (\Delta A_t) \oint_S k_{(0, -1)} $, as described in \cite{Compere:2007vx}. Therefore, one needs a compensating shift of $ \Phi_{int} $ to preserve the symmetry generator which defines the mass.

Therefore, it is natural to define the frame independent thermodynamical potentials: 
\begin{align}
T&=\alpha T_H,\\
\Omega&=\alpha(\Omega_H-\Omega_{int}),\\
\Phi&=\alpha(\Phi_H-\Phi_{int}).
\end{align}

Then, the first law \eqref{first law} takes the standard textbook form:
\bea \label{first law2}
\delta \mathcal{M} = T \delta S + \Omega \delta J + \Phi \delta Q.
\eea
We then recognize the thermodynamical quantities $T$, $\Omega$ and $\Phi$ as the chemical potentials associated to $S$, $J$ and $Q$, respectively. 
This expression contradicts the one presented in \cite{Gibbons:2013dna}, where there is an additional $\delta B$ term.

Another relation among the black hole conserved charges is the Smarr formula.
Let us review that the Smarr formula follows from the first law and Euler's theorem for homogeneous functions (see e.g. the excellent lecture notes \cite{Townsend:1997ku}).\footnote{Euler's theorem states that any homogeneous function of degree $n$ of $N$ variables, i.e., such that $f(t \bold{x})=t^n \bold{x}$, satisfies the relation 
$
\sum_{i=1}^N x^i \frac{\p f}{\p x^i}=n f(\bold{x}).
$}
The main observation is that the mass is a function of $S$, $J$, $Q^2$, i.e., $\mathcal{M}=\mathcal{M}(S, J, Q^2)$, and its variables have the same dimension, i.e., $[S]=[J]=[Q^2]=[mass^2]$ (in geometrized units $G=c=1$). Therefore, the mass $\mathcal{M}$ must be homogeneous of degree $n=\frac{1}{2}$ and it fulfils the relation
\bea
 \frac{1}{2}\mathcal{M}= \frac{\p \mathcal{M}}{\p S} S + \frac{\p \mathcal{M}}{\p J} J +  \frac{\p \mathcal{M}}{\p Q^2} Q^2.
\eea
After using $\frac{\p \mathcal{M}}{\p Q^2} Q^2 = \frac{1}{2} \frac{\p \mathcal{M}}{\p Q} Q$ and the first law, we get the Smarr relation:
\bea \label{Smarr}
\mathcal{M}= 2T S + 2\Omega J + \Phi  Q .
\eea

\section{Definition of the mass from integrability} \label{sec 3}

Let us first obtain the angular momentum $J$ and electric charge $Q$. By definition, the angular momentum and electric charge are the conserved charges associated with the symmetry parameters $(-\partial_{\phi}, 0)$ and $(0, -1)$, respectively. Using the Barnich-Brandt method \cite{Barnich:2001jy}, we directly note that the infinitesimal charges $\delta Q$ and $\delta J$ obey the integrability conditions \eqref{intcond}. Using the definition \eqref{defQ}, we obtain 
\begin{align}
J & \equiv \mathcal{Q}_{(-\partial_{\phi},0)} = j - q^3 B - \frac{3}{2} j q^2 B^2 - \frac{1}{4} q (8 j^2 + q^4) B^3 - \frac{1}{16} j (16 j^2 + 3 q^4) B^4 ,\label{ch2} \\ \label{charge}
Q & \equiv \mathcal{Q}_{(0, -1)} = q + 2 j B - \frac{1}{4} q^3 B^2,
\end{align}
where we defined $j = a m$ for convenience.
Both the angular momentum and the total electric charge match with the results of the literature: see e.g. \cite{Gibbons:2013dna, Astorino:2015naa, Booth:2015nwa2}.

\subsection{Integrability condition}

Let us now compute the infinitesimal change of energy caused by a change of all parameters of the solution $(\delta m, \delta j,\delta q,\delta B)$. Since it is not clear which canonical generator is associated with the conserved energy, let us first obtain the infinitesimal conserved charge $\slashed{\delta} \mathcal{Q}_{(\partial_{t},0)} = \oint_S k_{(\partial_{t},0)}$ associated with the symmetry parameter $(\partial_{t}, 0)$. Here $S$ is a sphere of integration around the black hole, e.g. $t,r$ constant. Using the definition of the surface charge and after a lengthly algebra, we obtain 
\bea \label{nonint}
\slashed{\delta} \mathcal{Q}_{(\partial_{t},0)} = c_{m}\delta m + c_{j}\delta j + c_{q}\delta q + c_{B}\delta B ,
\eea
where
\begin{align} \label{condef}
c_{m} &= 1 + \frac{3}{2} q^2 B^2 + 2j q B^3 + \Bigl(j^2 + \frac{q^4}{16}\Bigr)B^4,\\
c_{j} &= - 5 m q B^3 - \frac{1}{8}\frac{j}{m}(28m^2 + q^2)B^4,\\
c_{q} &= -3 m q B^2 - \frac{1}{4}\frac{j}{m}(8m^2 + 5q^2)B^3,\\
c_{B} &= \frac{j}{m}q + \frac{3}{2}m q^2 B - 3 j q \Bigl(2m + \frac{q^2}{4m^2}\Bigr)B^2 - \frac{1}{2}\Bigl(\frac{5}{4}mq^4 + \frac{j^2}{m}(8m^2+q^2)\Bigr)B^4.
\end{align}
We have used the symbol $\slashed{\delta}$ to emphasize that the expression \eqref{nonint} is not integrable in the parameter space. Indeed, it is easy to check that $\delta_1(\slashed{\delta}_2 \mathcal{Q}_{(\partial_{t}, 0)}) - (1\leftrightarrow 2)\neq 0$. Therefore, $\p_t$ is not associated with the energy.

Note that the coefficient $c_{m}$ exactly matches with the factor $\Xi$ \eqref{xi} introduced as a modification of the periodicity of the $\phi$ angle after the generation of the magnetic field parameter $B$ using the Ernst-Wild generating technique in order to avoid conical defects in  \cite{Astorino:2015naa, Gibbons:2013yq}.

The goal of this section is now to define the mass from integration over the phase space of an integrable infinitesimal canonical charge $\delta \mathcal{M}$ and prove that the procedure is unique and, therefore, that the mass $\mathcal{M}$ is uniquely defined. The black hole admits exactly three generalized Killing vectors and the symmetry associated with the mass can be any combination thereof. We can parametrize it as $\alpha(\chi, \lambda) = (\alpha (\partial_t + \Omega_{int}\partial_{\phi}), \alpha\Phi_{int})$.

We have to determine the four following functions defined over the parameter space: the mass $\mathcal{M}=\mathcal{M}(m, j, q,B)$ and the three constants which fix the canonical generator: $\alpha=\alpha(m, j, q,B)$, $\Omega_{int}=\Omega_{int}(m, j, q,B)$ and $\Phi_{int}=\Phi_{int}(m, j, q,B)$. These four functions obey the following equality by definition:
\bea
\delta \mathcal{M} = \alpha(\slashed{\delta} \mathcal{Q}_{(\partial_{t}, 0)} - \Omega_{int}\delta J - \Phi_{int}\delta Q). 
\eea
Quite remarkably, these defining equations are four equations since there are four parameters to be varied: $m,j,q,B$. They read in detail as 
\begin{align}
\partial_m \mathcal{M} &= \alpha \Xi,\label{int1}\\
\partial_q \mathcal{M} &= \alpha \Bigl(c_q - \partial_q Q \Phi_{int} - \partial_q J \Omega_{int}\Bigr),\label{int2}\\
\partial_B \mathcal{M} &= \alpha \Bigl(c_B - \partial_B Q \Phi_{int} - \partial_B J \Omega_{int}\Bigr),\label{consB}\\
\partial_j \mathcal{M} &= \alpha \Bigl(c_j - \partial_j Q \Phi_{int} - \partial_j J \Omega_{int}\Bigr).\label{int3}
\end{align}
From the first equation, we algebraically solve for $\alpha$ and obtain
\bea \label{alpha}
\alpha=\frac{1}{\Xi}\partial_m \mathcal{M}.
\eea
By algebraically solving the second and third equation for $\Phi_{int}$ and $\Omega_{int}$, we get
\begin{align} \label{Omega int}
\Omega_{int} &= \frac{1}{\alpha (\partial_q J \partial_B Q - \partial_B J \partial_q Q )}\Bigl[\alpha(c_q \partial_B Q - c_B \partial_q Q ) + \partial_q Q \partial_B \mathcal{M} - \partial_B Q \partial_q \mathcal{M}\Bigr], \\ 
 \label{Phi int}
\Phi_{int} &= \frac{1}{\alpha (\partial_q J \partial_B Q - \partial_B J \partial_q Q )}\Bigl[\alpha(c_B \partial_q J - c_q \partial_B J ) + \partial_B J \partial_q \mathcal{M} - \partial_q J \partial_B \mathcal{M}\Bigr]. 
\end{align}
One can check that the denominator $\partial_q J \partial_B Q - \partial_B J \partial_q Q$ only vanishes when the charge parameter $q=0$ or $q=3jB$. However both $\Omega_{int}$ and $\Phi_{int} $ shall be well defined, except for the trivial case $m=0$, upon substituting $\alpha$ and the mass $\mathcal{M}$ into their expressions.

Finally, by substituting $\alpha$, $\Omega_{int}$ and $\Phi_{int}$ in the last equation, we obtain a first-order linear homogeneous partial differential equation for $ \mathcal{M}$:
\bea\label{master}
\boxed{
\bigl[2 q (2 j+ 3 q m^2 B)\partial_m + 2 m (4 j + 3 q^3 B)\partial_q  - 4 m q^2( q - 3 j B)\partial_j- m (4 + 9 q^2 B^2)\partial_B\bigr]\mathcal{M}= 0.}
\eea

\subsection{Solution to the integrability condition}

We reduced the integrability requirement for defining the mass to a single partial differential equation for the mass \eqref{master}. Let us now solve it. First note that $B = 0$ is a regular point of the differential operator. Therefore, one has an analytic solution around $B=0$, 
\bea \label{ansatz}
\mathcal{M}(m, j, q, B) = \sum_{n \geq 0} f_{n}(m, j ,q) B^n.
\eea
The differential equation requires a boundary condition at $B = 0$. We now impose the physical requirement that the mass should coincide with the Kerr-Newman mass $m$ in the absence of a magnetic field, $B = 0$. Thus, we set $f_{0}(m, j, q) = m$.  It can be shown that the ansatz \eqref{ansatz} leads to an infinite series in the $B$ expansion which we can solve exactly to each order in $B$. 

It turns out to be simpler to consider the expansion of the mass squared: 
\bea \label{ansatz2}
\mathcal{M}^{2}(m, j, q, B) =m^{2}+\sum_{n \geq 1} g_{n}(m, j ,q) B^n
\eea
The functions $g_{n}(m, j ,q)$ must satisfy the following differential equation:
\bea \label{coeff}
m (4 + 9 q^2 B^2) \sum_{n \geq 1} n g_{n}(m, j ,q) B^{n-1} - \sum_{n \geq 1} \mathscr{D}[g_{n}(m, j ,q)]B^{n} - 4 m q (2 j+ 3 q m^2 B)=0,
\eea
where we defined the operator $\mathscr{D} = 2 q (2 j+ 3 q m^2 B)\partial_m + 2 m (4 j + 3 q^3 B)\partial_q  - 4 m q^2( q - 3 j B)\partial_j$. 

By collecting the terms order by order in the above equation, we get
\begin{align}
g_{1} &=2jq,\\
g_{2} &=\frac{1}{2m}\Bigl(3 m^{3}q^{2} +\mathscr{F}[g_{1}]\Bigr)=2 j^{2} + \frac{3}{2}m^{2}q^{2}-q^{4},\\
g_{n}&=\frac{1}{4 n m}\Bigl[3mq^{2}\bigl(-3(n-2)g_{n-2}+ 2\mathscr{G}[g_{n-2}]\bigr)+4\mathscr{F}[g_{n-1}]\Bigr], \label{coeff2}
\end{align}
where the differential operators $\mathscr{F}$ and $\mathscr{G}$ are defined as
\begin{align}
\mathscr{F}&=q j \partial_{m} - m q^{3}\partial_{j} +2m j \partial_{q},\\
\mathscr{G}&=m \partial_{m}+2j \partial_{j}+q \partial_{q}.
\end{align}
We would like to stress that the coefficients are uniquely determined, so that the mass $\mathcal{M}$ is unique. 
We can solve explicitly for the coefficients $g_{n}$. We observe that $g_{n}=0  \quad \forall n\geq 5$. Thus, the solution \eqref{ansatz2} has the remarkable advantage of admitting a finite $B$ expansion:
\begin{empheq}[box=\fbox]{align} \label{MIH}
\mathcal{M}^{2}(m, j, q, B)&= m^{2} + 2 j q B + \Bigl(2 j^{2} + \frac{3}{2}m^{2}q^{2}-q^{4}\Bigr)B^{2} + \\
	\qquad &+ j q\Bigl(2 m^{2} - \frac{3}{2} q^{2}\Bigr)B^{3}+\Bigl(j^{2}m^{2} - \frac{1}{2} j^{2} q^{2} + \frac{1}{16} m^{2} q^{4}\Bigr)B^{4}. \nonumber 
\end{empheq}
This is, therefore, the unique mass of the Kerr-Newman black hole immersed in a backreacting external magnetic field. This answer agrees with the mass computed in \cite{Booth:2015nwa2} by using the isolated-horizon approach. It disagrees with the other proposals in the literature.

% Besides the mass \eqref{MIH} computed using the covariant phase space and integrability methods \cite{Barnich:2001jy, Barnich:2007bf}, so far, there has been several different proposals for the mass of magnetised Kerr-Newman black hole, for instance see \cite{dokuchaev1987black},  \cite{aliev1988effective}, \cite{aliev1989magnetized}, \cite{karas1991interpretation} \cite{Gibbons:2013dna}, \cite{Astorino:2015naa} and \cite{Booth:2015nwa2}. The only result whose mass is integrable and that reproduce the extremal mass formula \eqref{mass extremal} was computed in \cite{Booth:2015nwa2} by means of the isolated-horizon technique. It exactly coincides with our result Eq. \eqref{MIH}.

\section{Thermodynamics}
 \label{sec 4}

In this section, we first give some properties of the mass. We then turn to the definition of the remaining thermodynamical potentials and the definition of the canonical frame of the Melvin-Kerr-Newman black hole. 

\subsection{Properties of the mass}

Extremality is reached for the Melvin-Kerr-Newman black hole when the mass parameter is given by $m=\sqrt{a^{2}+q^{2}}$. If the extremality condition is fulfilled, Eq. \eqref{MIH} is a perfect square and the extremal mass is given by
\bea\label{mass extremal}
\mathcal{M}_{ext}(a, q, B)=\sqrt{a^{2}+q^{2}}+a q B + \frac{1}{4}\sqrt{a^{2}+q^{2}}(4a^{2}+q^{2})B^{2}.
\eea
The same result was obtained in \cite{Astorino:2015naa} and \cite{Bicak:2015lxa} by the following independent approach. It is known that the near-horizon geometries of the extremal Kerr-Newman and the extremal magnetized Kerr-Newman black holes can be mapped to each other by a proper rescaling of the black hole parameters \cite{Astorino:2015naa,Bicak:2015lxa,Siahaan:2015xia}. Such a map correctly reproduces the angular momentum and the total electric charge of the magnetized Kerr-Newman black hole. Thus, even though the mass of the magnetized Kerr-Newman black hole was still unknown, the rescaling map was used to infer the mass of the extremal magnetized Kerr-Newman black hole from the mass of the extremal Kerr-Newman black hole.

%\begin{itemize}
%\item[1.] that computed in \cite{Gibbons:2013dna} by means of Kaluza-Klein reduction equals to $\mathcal{M}_{KK} = m \Xi$,
%	which does not obey the master equation \eqref{master}, so it is not integrable, and it does not reproduce the extremal mass formula \eqref{mass extremal};
%\item[2.] that computed in \cite{Booth:2015nwa} by means of the isolated-%horizon technique, which is exactly our result Eq. \eqref{MIH}.
%\end{itemize}

We conclude this section with an interesting observation about the mass \eqref{MIH}. It is nothing but the Christodoulou-Ruffini mass originally derived for the Kerr-Newman black hole \cite{Christodoulou:1972kt} given by
\bea \label{CR mass}
\boxed{
\mathcal{M}^{2}(S, J, Q)=\frac{S}{4\pi} + \frac{Q^{2}}{2} +\frac{\pi (Q^4+4J^2)}{4 S}.
}
\eea
When expressed in terms of extensive quantities, it clearly does not depend explicitly on the external magnetic field $B$. This result is in agreement with the general proof of the first law presented in Sec. \ref{sec 2}, where no additional $\delta B$ term is present. By retrospect, the first law \eqref{fl} and the existence of a perturbative series in $B$ of all physical quantities implies the mass formula \eqref{CR mass} since this formula holds in the absence of the magnetic field and the magnetic field cannot appear explicitly when it is turned on. 

By solving for the entropy $S$ in Eq. \eqref{CR mass} and by imposing it to be real, we obtain that 
\bea
0 \leq \mathcal{M}^{4} -Q^2\mathcal{M}^{2} - J^2 = m^2(m^2-a^2-q^2)\Xi^2,
\eea
from which we get the following lower bound of the mass $\mathcal{M}$ in terms of $Q$ and $J$:
\bea \label{bound}
\mathcal{M}^{2} \geq \frac{Q^2 + \sqrt{Q^4+4J^2}}{2}. 
\eea
This condition is equivalent to the absence of naked singularities, which in terms of black hole metric parameters reads as $m^2 \geq a^2 + q^2$. The bound is saturated at extremality. After transforming the inequality \eqref{bound} into an equality, the identity \eqref{CR mass} becomes
\bea \label{Sextr}
S_{ext}=\pi \sqrt{Q_{ext}^4+4J_{ext}^2}
\eea
in terms of physical charges at extremality.

Let us study more closely black holes without net electric charge $Q=0$. For vanishing external magnetic field $B=0$, the uncharged black hole condition trivially implies that the charge parameter $q=0$ and thus we recover the standard Kerr black hole described by $\mathcal{M}=m$ and $J=j$. In the presence of nonvanishing magnetic field $B \neq 0$, the constraint $ Q = 0$ can be most easily solved from Eq. \eqref{charge} by expressing $a=j/m$ in terms of the other three parameters $m,q,B$ as 
\bea \label{a_uncharged}
a = -\frac{q}{2 B m}\left(1 -\frac{B^2 q^2}{4} \right).
\eea
In this case, the bound \eqref{bound} becomes $\mathcal{M}^2 \geq |J|$. The bound is obeyed for either (i) $q=0$ which implies $a=0$ and in turn $\mathcal{M}=m$, $J=0$, (ii) $qB=\pm 2$ which implies $a=0$ and in turn $\mathcal{M}^2=4(2m^2-q^2)$, $J = \mp 4q^2$ or (iii) $m \geq |q|$ and either $B_{+} \leq B \leq -B_{-}$ or $B_{-} \leq B \leq -B_{+}$ with
\bea
B_{\pm}= -\frac{4m \sqrt{m^2-q^2}}{|q|^3} \pm \Bigl(\frac{4m^2}{|q|^3} -\frac{2}{|q|} \Bigr).
\eea
Notice that $B_{-} + B_{+}=-8m \sqrt{m^2-q^2}/|q|^3 \leq 0$. The case (i) describes the Melvin-Schwarzschild black hole, while cases (ii) and (iii) describe the Melvin-Kerr black hole.

Extremality amounts to imposing $m = \sqrt{a^2+q^2} $. There is no extremal limit in case (i), because otherwise we get $m=0$. The second case (ii) admits an extremal limit but there is no smooth solution in the limit $B \rightarrow 0$. In case (iii), one solves the equation $m = \sqrt{a^2+q^2} $ for $B$ and ones finds two distinct extremal branches $B=-\mbox{sign}(aq)B_{\pm}$. One finds
\bea
S_{ext} = 2\pi |J_{ext} |= 2 \pi \mathcal{M}_{ext}^2,\qquad |J_{ext}|=\frac{1}{2}\Xi_{ext} (2m^2-q^2).
\eea
where $\Xi_{ext}$ is computed from Eq. \eqref{xi}. The only free parameters are $m$ and $q$. The branch $B=-\text{sign}(aq)B_+$ admits a smooth limit to $B=0$ when $q \rightarrow 0$. One then recovers the extremal Kerr black hole with $\mathcal M_{ext} = m$, $|\mathcal J_{ext}| = m^2$.

\subsection{Thermodynamic potentials}

With the unique mass at hand, we obtain the expression of $\alpha$ from equation \eqref{alpha}. It is given by
\begin{align}
\alpha&=\frac{m}{\mathcal{M}}. 
\end{align}

We can then derive $\Omega_{int}$ and $\Phi_{int}$ from equations \eqref{Omega int} and \eqref{Phi int}. Their expressions are well defined except for the trivial case $m=0$. When the external magnetic field is turned off, we can check that we get $\alpha=1$, $\Omega_{int}=0=\Phi_{int}$ as expected in an asymptotically flat spacetime.

Using the results of Appendix A and the results above, the thermodynamical quantities defined in Sec. \ref{sec 2} are then explicitly given by
\begin{align} \label{MKN thermo}
T&=\alpha T_H = \Xi \frac{m}{\mathcal{M}} \frac{2}{A_H}(r_+ - m),\\
\Omega&=\alpha(\Omega_H-\Omega_{int})=\frac{1}{\Xi}\frac{J}{\mathcal{M}}\frac{1}{r_+^2+a^2},\\
\Phi&=\alpha(\Phi_H-\Phi_{int})=\frac{1}{\Xi}\frac{m}{\mathcal{M}}Q\Bigl[\frac{r_+}{r_+^2+a^2}\Bigl(\frac{Q}{q}\Bigr)^2+\frac{m}{ q^2}\Bigl( \Xi -\frac{\mathcal M^2}{m^2}\Bigr)\Bigr].
\end{align}
The quantities $r_+$, $T_H$, $\Omega_H$, $\Phi_H$, $A_H$ are defined in Appendix \ref{App}. 
For $B=0$, we recover the well-known expressions of the Kerr-Newman black hole. We can also check that the thermodynamic potentials coincide with the ones derived from the fundamental relation \eqref{CR mass}:
\bea
T &=& \frac{\p \mathcal M}{\p S} = \frac{1}{8 \pi \mathcal M} \left[1- \frac{4 \pi^2}{S^2} \left(J^2+\frac{Q^4}{4}\right) \right],\\
 \Omega &=& \frac{\p \mathcal M}{\p J} = \frac{\pi J}{\mathcal M S}, \\
  \Phi &=& \frac{\p \mathcal M}{\p Q} = \frac{Q}{2 \mathcal M S}(S+\pi Q^2).
\eea
By construction, the first law \eqref{first law2} and the Smarr relation \eqref{Smarr} are verified. 
%The thermodynamic stability can be analysed along the lines of \cite{Davies:1978mf}. 

The external magnetic field parameter $B$ only appears implicitly in the thermodynamic quantities $T$, $\Omega$ and $\Phi$. That means that the study of the thermodynamic stability against thermal or electric fluctuations is unchanged with respect to the Kerr-Newman black hole \cite{Davies:1978mf}. For example, the expressions for the heat capacity and electric permittivity are identical as in the Kerr-Newman case in terms of explicit thermodynamic variables.

\subsection{The canonical frame} \label{sec 5}
In this last paragraph, we show how to go to the canonical frame, namely the only frame in which the mass $\mathcal{M}$ is associated with the canonical symmetry generator $(\partial_{t_{can}},0)$.

We have computed the mass associated with the symmetry parameters $ \alpha(\partial_t + \Omega_{int}\partial_{\phi}, \Phi_{int})$.
We want to find a change of coordinates and a gauge transformation such that
\bea
\alpha(\partial_t + \Omega_{int}\partial_{\phi}, \Phi_{int}) \rightarrow (\partial_{t_{can}},0).
\eea
According to \eqref{cc}, the change of coordinates is given by
\begin{align}
t &\rightarrow t_{can}=\frac{t}{\alpha};\\
\phi&\rightarrow \phi_{can}=\phi - \Omega_{int} t
\end{align}
while the large gauge transformation of the potential $A^{can}_{\mu}=A_\mu+\partial_{\mu}\lambda$ is given by $\lambda = \alpha \Phi_{int}$. The total transformation of $A_t$ is, therefore,
\bea
A_t \rightarrow A^{can}_{t}=\alpha( A_t + \Omega_{int} A_\phi + \Phi_{int} ) .
\eea

\subsection{Alternative thermodynamics: Magnetic field as a source}

If one allows for all physical quantities ($\mathcal M, Q, J ,B$) to be varied, the definition of mass is unique as we have shown. It matches with the mass computed using the isolated horizon formalism \cite{Booth:2015nwa2} and at extremality using the match with the near-horizon of the Kerr-Newman black hole \cite{Astorino:2015naa,Bicak:2015lxa}. This existence and uniqueness result suggests that the solution space of Melvin-Kerr-Newman metrics with arbitrary freely varying parameters constitutes a phase space with well-defined boundary conditions. 

Now, alternative boundary conditions can lead to alternative definitions of the gravitational mass, as illustrated e.g. in the case of gravity coupled to scalar fields \cite{Henneaux:2006hk} with bare mass in the Breitenlohner-Freedman range  which allows for various boundary conditions \cite{Breitenlohner:1982bm} (see also related cases in supergravity \cite{Lu:2013ura,Chow:2013gba,Lu:2014maa,Anabalon:2015xvl}). In particular, other boundary conditions could be formulated if one instead considers the magnetic field $B$ as a fixed source. In this case, no integrability condition needs to be imposed for variations of the magnetic field, and the constraint \eqref{consB} should not be imposed. This leads to additional possibilities for defining the mass. 

More precisely, one defines new quantities $ \widetilde \alpha,\, \widetilde \Omega_{int}$ and $ \widetilde \Phi_{int}$. The canonical infinitesimal charge $\slashed{\delta} \mathcal{M}$ associated with the generator $ \widetilde \alpha(\p_t +  \widetilde \Omega_{int} \p_\phi,  \widetilde \Phi_{int})$ is not integrable for arbitrary variations of the magnetic field $B$ but it can be written as
\bea
\slashed{\delta} \mathcal{M} = \delta \widetilde{\mathcal M} + \mu \delta B \label{relM}
\eea
The left-hand side of \eqref{relM} is exactly the canonical charge which can be deduced from \eqref{ch2}-\eqref{charge}-\eqref{nonint}. Once $\widetilde{\mathcal M}=\widetilde{\mathcal M}(m,a,q,B)$ is fixed as a function of the parameters, $\mu$ is uniquely defined from \eqref{relM}. There are four unknown functions $ \widetilde \alpha, \widetilde \Omega_{int}, \widetilde \Phi_{int},\widetilde{\mathcal M}$ and three integrability conditions \eqref{int1}-\eqref{int2}-\eqref{int3}. There is, therefore, one free function, the mass, which should be fixed by independent considerations such as a boundary condition. A physical requirement is that the mass reduces to the Kerr mass in the absence of magnetic field, $\widetilde {\mathcal M} = m + O(B)$.  Otherwise, the mass is arbitrary with this method. Once the remaining function is fixed, one can compute $\mu$ from \eqref{relM} which can be interpreted as a conjugate chemical potential for the magnetic field, i.e. an induced magnetic dipole moment. Indeed, after using \eqref{relM}, the first law \eqref{first law2} reads as 
\bea \label{nf}
 \delta \widetilde{\mathcal M} + \mu \delta B  = \widetilde T \delta S + \widetilde \Omega \delta J +\widetilde \Phi \delta Q,
\eea
where the tilded chemical potentials are defined as
\begin{align}
 \widetilde T&= \widetilde \alpha T_H,\\
 \widetilde \Omega&= \widetilde \alpha(\Omega_H- \widetilde\Omega_{int}),\\
 \widetilde \Phi&=\widetilde \alpha(\Phi_H- \widetilde\Phi_{int}).
\end{align}
Since $B$ has mass dimension $[mass]^{-1}$, the Smarr relation reads as 
\bea
\widetilde {\mathcal M}= 2 \widetilde  T_H S + 2\widetilde{\Omega} J + \widetilde{\Phi}  Q + \mu B. 
\eea
As an intermediate summary,  one can arbitrarily define the mass with the integrability procedure, up to the constraint of matching with $m$ in the absence of magnetic field, and still get a consistent thermodynamics with first law and Smarr formula. 

One such definition of mass was recently given in \cite{Gibbons:2013dna} after using a Kaluza-Klein reduction to three dimensions. Its expression is given by
\bea
\mathcal M_{GPP}= \Xi m,\label{mass}
\eea
where the factor $\Xi$ defined in \eqref{xi} naturally appears in the regulation of the metric in order to avoid conical defects. Identifying $\widetilde {\mathcal{M}} = \mathcal M_{GPP}$ one can solve the three integrability conditions \eqref{int1}-\eqref{int2}-\eqref{int3} with 
\begin{align}
\widetilde{\alpha} &=1\\
\widetilde{\Omega}_{int} &= \frac{B^3\bigl[160 q m^2 + j (80m^2+76q^2)B +8q(4j^2+31m^2q^2)B^2 + 3j q^2 (44m^{2}+q^{2})B^3\bigr]}{8\Xi m(4+9q^2B^2)}\\
\widetilde{\Phi}_{int} &= \frac{1}{16\Xi m (4+9q^2B^2)}\Bigl[ -384 q m^2B^2 - 16j(16m^2+5q^2)B^3 - 16q(2j^2+53m^2q^2)B^4 \nonumber\\  & +96jq^2(q^2-5m^2)B^5 + 8(-46m^2q^5 +qj^2(28m^2+43q^2))B^6 + \\
& + j(-408m^2q^4+5q^6+32j^2(2m^2+11q^2))B^7 +3q(32j^4-8j^2m^2q^2+5m^2q^6)B^8 \Bigr]. \nonumber
\end{align}
This leads from \eqref{relM} to the magnetic dipole moment:
\bea
\mu = \Xi  \frac{2q}{m} \frac{2j-3Bqm^2}{4+9B^2q^2}.
\eea
In the weak magnetic field approximation both $\widetilde{\Omega}_{int}$ and $\widetilde{\Phi}_{int}$ vanish, whereas $\mu$ approaches $\frac{jq}{m} \simeq \frac{JQ}{\mathcal M}$, which is the well-known magnetic dipole moment of a charged stationary axisymmetric vacuum black hole spacetime \cite{Wald:1974np}. 

In summary, the mass proposed in \cite{Gibbons:2013dna} leads to a consistent thermodynamics in the case where the magnetic field is considered as an external source.  The definition $\mathcal M_{GPP}$ however does not match with the extremal mass \eqref{mass extremal} computed from the near-horizon matching \cite{Astorino:2015naa,Bicak:2015lxa}. We found that in all generality, integrability methods are ambiguous up to one arbitrary function which can be precisely identified with the mass. There is, therefore, a one-function family of consistent thermodynamics when the magnetic field is considered as an external source. We leave open the problem whether or not a mass exists in this framework which matches with the Kerr mass in the absence of magnetic field and which also matches at extremality with the near-horizon extremal mass \cite{Astorino:2015naa,Bicak:2015lxa}. The relationship between the mass formula and consistent boundary conditions for Melvin-Kerr-Newman spacetimes is also left open.

\section*{Acknowledgments}

We thank Y. Pang and C. Troessaert for interesting comments. G.C. is Research Associate of the Fonds de la Recherche Scientifique F.R.S.-FNRS (Belgium). G.C. and R.O. both acknowledge the current support of the ERC Starting Grant No. 335146 ``HoloBHC". This work is also partially supported by FNRS-Belgium (convention IISN 4.4503.15). M. A. is funded by the Conicyt-PAI Grant No. 79150061. The Centro de Estudios Cient\'ificos (CECs) is funded by the Chilean Government through the Centers of Excellence Base Financing Program of Conicyt.

\appendix

\section{Black hole fields and physical quantities} \label{App}

In this appendix, we first describe the exact metric and gauge field which we use through the paper. We will also recall the thermodynamical quantities defined at the horizon of the magnetized Kerr-Newman black hole in the reference frame used in \cite{Gibbons:2013yq}. 

The black hole depends upon four parameters: the mass parameter $m$, the electric charge parameter $q$, the angular momentum parameter $a$ and the magnetic field $B$. The metric and gauge field were clearly written in a specific coordinate system and gauge in Appendix B of \cite{Gibbons:2013yq}: see their definition (B.4). Since these fields would be very long to rewrite, we will simply refer the reader to that reference. We will use this explicit form of the solution up to three modifications which we now explain. 

We first set the magnetic monopole charge parameter $p$ of \cite{Gibbons:2013yq} to zero in order to cancel the magnetic monopole charge. Second, we use a $\phi$ angle which is $2\pi$ periodic. It is related to the $\phi_{GMP}$ angle defined in \cite{Gibbons:2013yq} by $\phi_{GMP} = \Xi \,  \phi $ where the constant $ \Xi $ was firstly computed in \cite{hiscock1981black} in order to ensure that the metric does not admit conical defects. It reads, explicitly,
\bea
 \Xi = 1 + \frac{3}{2} q^2 B^2 + 2j q B^3 + \Bigl(j^2 + \frac{q^4}{16}\Bigr)B^4.\label{xi}
\eea
Finally, we added the constant  $A^{(0)}_{\phi}$ to the gauge potential component $A_{\phi}$ in order to ensure regularity of the gauge potential along the axis of rotation. Namely, $A_{\phi}=0$ at the North and South poles. This leads us to set
\bea
A^{(0)}_{\phi} =- \Bigl( \frac{3}{2} q^2 B + 3 j q B^2 + \frac{1}{8}(q^4+16 j^2) B^3 \Bigr).
\eea

Let us now review the thermodynamical quantities defined at the horizon of the magnetized Kerr-Newman black hole. The outer and inner horizons are located at $r_{\pm}=m\pm \sqrt{m^2-a^2-q^2}$. These radial coordinate values are not affected by the presence of the external magnetic field $B$. In the given coordinate system $(t,r,\theta,\phi)$ and gauge,  the angular velocity evaluated at the event horizon $r_+$ is given by
\bea
\Omega_H=-\frac{g_{t\phi}}{g_{\phi\phi}}\Bigg|_{r=r_{+}}=\frac{\omega^{(GMP)}_H}{ \Xi },
\eea
where $\omega^{(GMP)}_H$ is the function defined in (B.8) of \cite{Gibbons:2013yq} evaluated at $r=r_{+}$.
The Killing generator of the black hole horizon is $\xi=\partial_t+\Omega_H\partial_{\phi}$. The Coulomb electrostatic potential is
\bea
\Phi_H=- A_{\mu} \xi^{\mu} |_{r=r_{+}}=-\Omega_H A_{\phi}^{(0)} - \Phi^{(GMP)}_0,
\eea
where $\Phi^{(GMP)}_0$ is the function defined in (B.17) of \cite{Gibbons:2013yq} evaluated at $r=r_{+}$. 
The surface gravity is given by
\bea
\kappa=\frac{r_+-r_-}{2(r^2_+ +a^2)},
\eea
and the Hawking temperature is given by $T_H=\frac{\kappa}{2\pi}$. Note that both the surface gravity $\kappa$ and the Hawking temperature $T_H$ do not depend on the parameter $B$ but only on $m,a,q$.
The black hole area is given by
\bea
A_H=4 \pi  \Xi  (r^2_+ + a^2),
\eea
and the entropy is $S=A_H/4$ in units where Newton's constant is $G=1$.

%\bibliography{refs}

%\providecommand{\href}[2]{#2}\begingroup\raggedright
%\endgroup

\end{document}